\documentstyle[12pt,epsf,overcite]{article}


\begin{document}

\baselineskip 14pt

 \rightline{SPhT-94-155}
\vspace{24pt}

\begin{center}
{\large\bf
The Yangian Deformation of the $W$-Algebras\par
and the Calogero-Sutherland System.
\footnote{To appear in the proceedings of the 6th Nankai workshop.}
}

\vspace{24pt}
Denis {\sc Bernard}\footnote{Member of the C.N.R.S.},
Kazuhiro {\sc Hikami}$^\dagger$,
and
Miki {\sc Wadati}$^\dagger$

\vspace{10pt}
{\sl Service de Physique Theorique de Saclay,} \\
{\sl F-91191, Gif-sur-Yvette, France.}

\vspace{8pt}

$^\dagger$
{\sl Department of Physics, Faculty of Science,} \\
{\sl University of Tokyo,} \\
{\sl Hongo 7-3-1, Bunkyo, Tokyo 113, Japan.}

\vspace{8pt}

\end{center}

\vspace{18pt}

\begin{center}
ABSTRACT
\end{center}
\hspace{0.5in}
\begin{minipage}[t]{5in}
{\small

The Yangian symmetry Y(su($n$)) of the
Calogero-Sutherland-Moser spin model is reconsidered.
The Yangian generators are constructed from two
non-commuting su($n$)-loop algebras.
The latters generate an infinite dimensional symmetry
algebra which is a deformation of the $W_\infty$-algebra.
We show that this deformed $W_\infty$-algebra contains
 an infinite number of Yangian
subalgebras with different deformation parameters.
}
\end{minipage}



\vspace{1cm}
\par\noindent
{\bf Introduction.}

We consider the Calogero and the Sutherland spin models.
These models are  inverse square interacting  $ N $-body systems
with internal degree of freedom.
Their hamiltonians  are respectively defined as
\begin{eqnarray}
  {\cal H}_C & = & - \sum_{j=1}^N \frac{\partial^2}{\partial x_j^2} +
  \sum_{j \neq k}^N \frac{ \lambda^2 - \lambda \, P_{jk} }
  { (x_j - x_k)^2 } ,
  \label{hcal} \\
  {\cal H}_S & = & \sum_{j=1}^N
  \Bigl( x_j \frac{ \partial}{\partial x_j} \Bigr)^2
  + \sum_{j \neq k}^N
  ( -\lambda^2 + \lambda \, P_{jk} )
  \frac{ x_j \, x_k}{ (x_j - x_k)^2 } .
  \label{hsuth}
\end{eqnarray}
Here $ P_{jk} $ is the permutation operator in the su($n$) spin space.
Using the spin operators $ E^{ab} $ as basis of the su($n$) algebra,
$ E^{ab} \equiv | a \rangle \langle b | $ ($ a, b=1, \cdots,n $),
the operator $ P_{jk} $ can be written as
\begin{equation}
  P_{jk} = \sum_{a,b=1}^n E_j^{ab} E_k^{ba} .
\end{equation}
The hamiltonian~(\ref{hsuth}) reduces to a dynamical particle system
with periodic boundary condition when we change the variables,
$ x_j \to \exp( - 2 \pi {\rm i}  z_j /L )$.

Both the Calogero~(\ref{hcal}) and the Sutherland~(\ref{hsuth})
models are
integrable~\cite{WadaHika94a,HaHal92,HiWa93a,MinPol93a,BGHP93}.
The integrability can be checked by several methods: e.g.
the quantum Lax formalism and the Dunkl operator formalism.
Usually, the Calogero and the Sutherland spin models are
treated separately.
In this note we construct  an algebra in which
both models are naturally included~\cite{WadaHika94a}.
Our algebra can be generated from two non-commuting su($n$) loop
subalgebras. It has an infinite number of Yangian subalgebras.

\bigskip
\noindent
{\bf The Calogero  Model.}

Let us first consider the Calogero system.
For our purpose we define the following two operators:
\begin{eqnarray}
  J_0^{ab} & = & \sum_{j=1}^N E_j^{ab} , \\
  J_1^{ab} & = & \sum_j E_j^{ab} \frac{\partial}{\partial x_j} -
  \lambda \sum_{j \neq k} ( E_j E_k )^{ab} \frac{1}{x_j - x_k} .
\end{eqnarray}
Here we have used the conventional notations,
$ ( E_j E_k )^{ab} = \sum_{c=1}^n E_j^{ac} E_k^{cb} $.
The generators $ J_0^{ab} $ and $ J_1^{ab} $
satisfy the following relations:
\begin{eqnarray}
  & & [ J_0^{ab} , J_0^{cd} ]  =  \delta^{bc} J_0^{ad} - \delta^{da}
  J_0^{cb} ,
  \label{j0j0} \\
  & & \bigl[ J_0^{ab} , J_1^{cd} \bigr]  =  \delta^{bc} J_1^{ad} -
  \delta^{da} J_1^{cb} , \\
  & & \bigl[ J_0^{ab} , \bigl[ J_1^{cd} , J_1^{ef} \bigr] \bigr] -
  \bigl[ J_1^{ab} , \bigl[ J_0^{cd} , J_1^{ef} \bigr] \bigr] = 0 .
  \label{loopserre}
\end{eqnarray}
The third equation is known as the Serre relation for the loop
algebra. These relations imply that the higher generators
$ J_{n>1}^{ab} $, which  are defined recursively using the generator
$ J_1^{ab} $, form a representation of su($n$) loop algebra,
\begin{equation}
    [ J_n^{ab} , J_m^{cd} ]  =  \delta^{bc} J_{n+m}^{ad} -
    \delta^{da}  J_{n+m}^{cb} .
    \label{loop}
\end{equation}
Remark that the generators of the su($n$) loop algebra
$ J_n^{ab} $ are conserved operators for the Calogero spin
model~\cite{HiWa93e,BGHP93},
\begin{equation}
  [ {\cal H}_C , J_n^{ab} ] = 0 .
\end{equation}
 From this we conclude that the Calogero spin
model~(\ref{hcal}) is su($n$) loop invariant.

\bigskip
\noindent
{\bf The Sutherland  Model.}

Consider now the Sutherland spin model~(\ref{hsuth}).
It can be viewed as the Calogero spin model
with periodic boundary condition. It is
not invariant under the su($n$) loop algebra.
However, the Sutherland spin model is invariant under
a ``deformed'' su($n$) loop algebra, or
in a recent mathematical terminology, a Yangian algebra
Y(su($n$))~\cite{HHTBP92,BGHP93}.
The Yangian algebra was first defined by Drinfeld  as a
Hopf algebra accompanied with the Yang's rational solution of the
quantum Yang-Baxter equation~\cite{Dri86b}.
To see that the Sutherland spin model~(\ref{hsuth}) possesses the
Yangian symmetry, we  introduce two generators as~\cite{HHTBP92,BGHP93},
\begin{eqnarray}
  Q_0^{ab} & = & J_0^{ab} , \\
  Q_1^{ab} & = & \sum_j E_j^{ab}
  \Bigl(
  x_j \frac{\partial}{\partial x_j} + \frac{1}{2}
  \Bigr)
  - \frac{\lambda}{2} \sum_{j \neq k} ( E_j E_k )^{ab} \frac{x_j +
    x_k}{x_j - x_k} .
\end{eqnarray}
One then directly check that generators $ Q_0^{ab} $ and $ Q_1^{ab} $ are
conserved operators for the Sutherland spin model:
\begin{equation}
  [ Q_0^{ab} , {\cal H}_S ]
  = [ Q_1^{ab} , {\cal H}_S ] = 0.
\end{equation}
After a lengthy calculation we have the following commutation
relations:
\begin{eqnarray}
  & & [ J_0^{ab} , Q_1^{cd} ]
  =  \delta^{bc} \, Q_1^{ad} - \delta^{da} \,  Q_1^{cb} ,
  \label{q0q1} \\
  & &  [ J_0^{ab} , [ Q_1^{cd} , Q_1^{ef} ] ] - [ Q_1^{ab} ,
  [ J_0^{cd} , Q_1^{ef} ] ] \nonumber \\
  & & \mbox{ } = \frac{\lambda^2}{4}
  \Bigl(
    [ J_0^{ab} , [ ( J_0 J_0 )^{cd} , ( J_0 J_0 )^{ef} ] ]
    - [ ( J_0 J_0 )^{ab} , [ J_0^{cd} , ( J_0 J_0 )^{ef} ] ]
  \Bigr).
  \label{serre}
\end{eqnarray}
These relations together with equation (\ref{j0j0}) are the
defining relations of the Yangian Y(su($n$)).
The second equation (\ref{serre}) is called  the ``deformed'' Serre
relation.  It reduces to
the Serre relation~(\ref{loopserre}) for the loop algebra
when $ \lambda \rightarrow 0 $.
In this sense, the Yangian can be viewed as a ``deformed'' loop algebra.
The relations (\ref{j0j0}) and (\ref{q0q1}-\ref{serre}) show that
the generators $ Q_0^{ab} $ and $ Q_1^{ab} $ form a representation
of the Yangian algebra  Y(su($n$)).
Since the Yangian generators $ Q_n^{ab} $ commute with the
Sutherland spin hamiltonian (\ref{hsuth}),
this model has the Yangian symmetry Y(su($n$)).

\bigskip
\noindent
{\bf The Yangian Deformed $ W_\infty $ Algebra.}

To combine the loop algebra $ J_n^{ab} $ and the Yangian algebra
$ Q_n^{ab} $,
we introduce another set of generators  $ K_n^{ab} $~\cite{HiWa93e}:
\begin{equation}
  K_n^{ab} = \sum_{j=1}^N E_j^{ab} x_j^n .
\end{equation}
It is easy to see that the generators  $ K_n^{ab} $ represent
the su($n$) loop algebra,
\begin{equation}
    [ K_n^{ab} , K_m^{cd} ]  =  \delta^{bc} K_{n+m}^{ad} -
    \delta^{da}  K_{n+m}^{cb} ,
\end{equation}
All the $ K_n^{ab} $ can be defined recursively from the
two  lowest generators,
\begin{eqnarray}
  K_0^{ab} & = & J_0^{ab} , \\
  K_1^{ab} & = & \sum_j E_j^{ab} x_j .
\end{eqnarray}
By construction, they satisfy the relations
(\ref{j0j0}-\ref{loopserre})
with $J^{ab}_n$ replaced by $K^{ab}_n$.

Consider now the algebra generated by the elements
$ \{ J_0^{ab} , J_1^{ab} , K_1^{ab} \} $.
The Yangian current $ Q_1^{ab} $  appears from
an inter-relation formula  between these operators;
\begin{equation}
  [ J_1^{ab} , K_1^{cd} ] + [ K_1^{ab} , J_1^{cd} ] = 2
  \Bigl(
    \delta^{bc} \, Q_1^{ad} - \delta^{da} \, Q_1^{cb}
  \Bigr) .
  \label{jkq}
\end{equation}
Besides the relation (\ref{serre}) for $Q^{ab}_1$,
we also have the following Serre-like relations,
\begin{eqnarray}
  & & [ J_0^{ab} , [ J_1^{cd} + Q_1^{cd} , J_1^{ef} + Q_1^{ef} ]]
   - [ J_1^{ab} + Q_1^{ab} , [ J_0^{cd} , J_1^{ef} + Q_1^{ef} ]]
   = 0 ,
   \label{jjq1} \\
  & & [ J_0^{ab} , [ K_1^{cd} + Q_1^{cd} , K_1^{ef} + Q_1^{ef} ]]
   - [ K_1^{ab} + Q_1^{ab} , [ J_0^{cd} , K_1^{ef} + Q_1^{ef} ]]
   = 0 .
   \label{jjq2}
\end{eqnarray}
The relations (\ref{j0j0}-\ref{loopserre}),
(\ref{q0q1}-\ref{serre}) and (\ref{jkq}-\ref{jjq2}) possess
an interesting interpretation: consider the
generators $ Q_1^{ab}(x,y) $ defined by
\begin{equation}
  Q_1^{ab} (x,y) \equiv Q_1^{ab} + x \, J_1^{ab} + y \, K_1^{ab},
\end{equation}
for any complex numbers $x$ and $y$.
Then, all the previously written Serre relations can be
summarized into the following compact equations~:
\begin{eqnarray}
  & & [ J_0^{ab} ,  Q_1^{cd}(x,y) ]
  =  \delta^{bc} \, Q_1^{ad}(x,y) -
  \delta^{da}  \, Q_1^{cb} (x,y) , \\
  & & [ J_0^{ab} , [ Q_1^{cd}(x,y) , Q_1^{ef}(x,y) ] ]
  - [ Q_1^{ab}(x,y) , [ J_0^{cd} , Q_1^{ef}(x,y) ] ] \nonumber \\
  & & \mbox{ }
  = \frac{\lambda^2}{4}
  \Bigl(
  [ J_0^{ab} , [ ( J_0 J_0 )^{cd} , ( J_0 J_0 )^{ef} ] ]
  - [ ( J_0 J_0 )^{ab} , [ J_0^{cd} , ( J_0 J_0 )^{ef} ] ]
  \Bigr) .
\end{eqnarray}
In other words, the commutation relations  between the
generators $J^{ab}_n$ and $K^{ab}_n$ of the two loop subalgebras
are such that the generators $ Q_1^{ab}(x,y)$ form a representation
of the Yangian for any $x$ and $y$.
We thus have an infinite number of Yangian subalgebras
constructed from $ Q_1^{ab}(x,y) $, but they all have $\lambda$
as deformation  parameter.

The algebra generated by  $ \{ J_0^{ab} , J_1^{ab} , K_1^{ab} \} $ is
schematically drawn in the figure.
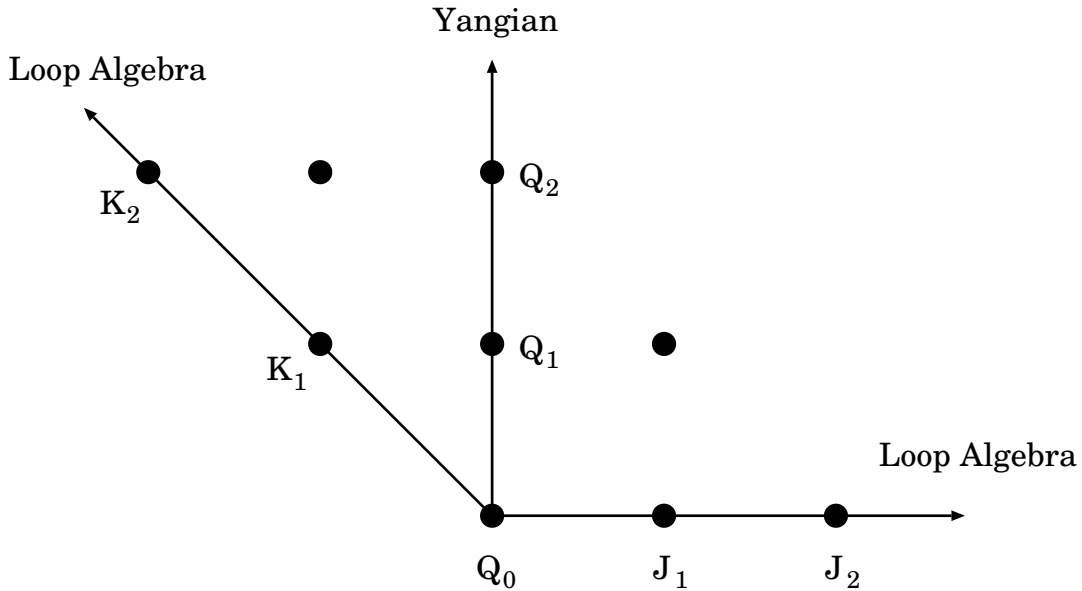
\begin{center}
  \quad{\def\epsfsize#1#2{0.9#1}\epsfbox{figcs.eps}}\quad
\end{center}
It is appropriate here to relate the notations in this note with those
in Ref.~\citen{HiWa93e};
the horizontal and vertical axes indicate the node $ n $
and the spin $ s $ respectively.
Corresponding to each solid circle, there exists an generator
$ Q_n^{(s)ab} $.  The infinite symmetry associated to
the operators $ Q_n^{(s)ab} $ was named
the quantum $ W_\infty $ algebra~\cite{HiWa93e,WadaHika94a}.

In the limit $\lambda\to 0$, the generators $J^{ab}_n$
reduce to $J^{ab}_n=\sum_j E^{ab}_j(\partial_{ x_j})^n$.
Together with the operators $K_n^{ab}$, they generate a
$W_\infty$-algebra with elements,
\begin{eqnarray}
  Q^{(s)ab}_{n}=\sum_{j=1}^N E^{ab}_j x_j^{s-1}
  ( \partial_{ x_j} )^{n+s-1} ,
\end{eqnarray}
which satisfy the commutation relations,
\begin{eqnarray}
  [  Q^{(s)ab}_{n} , Q^{(s')cd}_{m} ] & = &
  \delta^{bc} \cdot \sum_{k=0}^{n+s-1}
\frac{ (n+s-1)!(s'-1)!}{k!(n+s-k-1)!(s'-k-1)!}
  \, Q_{n+m}^{(s+s'-1-k)ad} \nonumber \\
  & & \ - \delta^{da}  \cdot \sum_{k=0}^{m+s'-1}
\frac{ (m+s'-1)!(s-1)!}{k!(m+s'-k-1)!(s-k-1)!}
  \, Q_{n+m}^{(s+s'-1-k)cb}  .  \nonumber \\
\end{eqnarray}
As a consequence, this algebra is generated by the elements
$ \{ J_0^{ab} , J_1^{ab} , K_1^{ab} \} $.
Moreover, it is easy to see that this $W_\infty$-algebra
possesses an infinite number of su($n$) loop subalgebras.

For $\lambda\not=0$, our algebra is naturally called
a ``Yangian deformed $W_\infty$-algebra", and denoted $YW_\infty(su(n))$.
The algebra includes the loop algebra, the Virasoro
algebra~\cite{HiWa93e}, and the Yangian as the subalgebras.

\bigskip
\noindent
{\bf The Yangian Subalgebras.}

We now analyze a little more the structure of the algebra.
Let us first identify another Yangian subalgebra.
Define another set of operators $ \widetilde{Q}_1^{ab} (h,\omega) $ by
\begin{equation}
  \widetilde{Q}_1^{ab} (h,\omega) =   h^2 J_2^{ab} - \omega^2 K_2^{ab}
  ,
\end{equation}
where $ h $ and $ \omega $ are arbitrary complex numbers.
By direct computation, we see that the operators
$ \widetilde{Q}_1^{ab} (h,\omega) $ constitute a representation of
the  Yangian since they satisfy the following relations:
\begin{eqnarray}
  & & [ J_0^{ab} , \widetilde{Q}_1^{cd}(h,\omega) ]
  =  \delta^{bc} \, \widetilde{Q}_1^{ad}(h,\omega) -
  \delta^{da} \, \widetilde{Q}_1^{cb}(h,\omega) , \\
  & & \bigl[ J_0^{ab} ,
  \bigl[
  \widetilde{Q}_1^{cd}(h,\omega), \widetilde{Q}_1^{ef}(h,\omega)
  \bigr]
  \bigr]
  - \bigl[ \widetilde{Q}_1^{ab}(h,\omega)  ,
  \bigl[ J_0^{cd} , \widetilde{Q}_1^{ef}(h,\omega) \bigr] \bigr] \nonumber
  \\
  & & \mbox{ } = (\lambda h\omega)^2
  \Bigl(
  [ J_0^{ab} , [ ( J_0 J_0 )^{cd} , ( J_0 J_0 )^{ef} ] ]
  - [ ( J_0 J_0 )^{ab} , [ J_0^{cd} , ( J_0 J_0 )^{ef} ] ]
  \Bigr).
\end{eqnarray}
Notice that the deformation parameter is now
$ 2\lambda h\omega $.
These Yangian generators are conserved operators  for the
Calogero spin model confined in a harmonic potential with hamiltonian,
\begin{equation}
  {\cal H}_{CM} = h^2 \, {\cal H}_C + \omega^2 \, \sum_j x_j^2 .
\end{equation}
Hence, the Calogero model~(\ref{hcal}) confined in
the harmonic potential also possesses
the Yangian symmetry~\cite{Hikam94b}.  The limit of $ \omega \to 0 $
corresponds to the Calogero spin model~(\ref{hcal}); in this case
the Yangian symmetry reduces to the loop algebra.

This subalgebra is actually a simple example of a more general
structure.
As we now explain, in the Yangian deformed $W_\infty$-algebra
generated by $\{J^{ab}_0,J^{ab}_1,K^{ab}_1\}$, there exists
an infinite number of ``slices" in which a Yangian subalgebra
can be constructed.

To prove it, we need to introduce the Dunkl operators $D_i$
for the Calogero model~\cite{Pol}.
\begin{eqnarray}
  D_i = {\partial \over \partial x_i} - \lambda
  \sum_{j:j\not=i} \frac{1}{x_i-x_j} K_{ij} .
  \label{dunkl}
\end{eqnarray}
where $K_{ij}$ is the operator permuting the coordinates $x_i$
and $x_j$:
$ x_i \, K_{ij}=K_{ij} \, x_j$.
We have the commutation relations:
\begin{eqnarray}
  D_i \, K_{ij} & = & K_{ij} \, D_j , \\
  \bigl[ D_i,D_j \bigr] & = & \bigl[x_i,x_j\bigr]=0,\\
  \bigl[ D_i,x_j \bigr] & = &
  \delta_{ij} \,
  ( 1+ \lambda \, \sum_{l:l \neq i} K_{il} )
  - (1 - \delta_{ij} ) \, \lambda \, K_{ij} .
\end{eqnarray}
Introduce now the operators $\Delta_i$ defined by:
\begin{equation}
\Delta_i = (h D_i + \omega x_i + y) \, (h' D_i + \omega' x_i + y').
\label{quadra}
\end{equation}
They depend on the c-numbers $h,\omega, y$ and $h',\omega',y'$.
They satisfy
\begin{equation}
  [ \Delta_i , \Delta_j ] = \lambda \, (h\omega'-h'\omega)
  \, ( \Delta_i- \Delta_j )
  \, K_{ij}. \label{comrel}
\end{equation}
This relation allows us to construct a representation
of the Yangian algebra. Following Ref.~\citen{BGHP93}, we introduce
a monodromy matrix $T(u)$ by
\begin{equation}
  T^{ab}(u)=\delta^{ab} + \lambda \, (h\omega'-h'\omega)
  \sum_i \pi
  \Bigl(
  \frac{1}{u-\Delta_i}
  \Bigr)
  \, E^{ab}_i ,
\label{monot}
\end{equation}
where  $\pi$ is the
projection consisting in replacing $K_{ij}$ by $P_{ij}$ once the
permutation $K_{ij}$ has been moved to the right of the expression.
The matrix $T^{ab}(u)$ satisfy,
\begin{equation}
  [ T^{ab}(u) , T^{cd}(v) ] =
  \frac{ \lambda(h\omega'-h'\omega) }{u-v}
  \Bigl(
  T^{cb}(u) \, T^{ad}(v) - T^{cb}(v) \, T^{ad}(u)
  \Bigr)
\end{equation}
This is another presentation of the Yangian. Therefore,
the matrix~(\ref{monot}) forms a representation of the
Yangian. As usual, the quantum determinant of $T(u)$
defines a generating function of commuting operators
which all commute with the matrix $T(v)$ itself.

We thus have identified an infinite number of Yangian subalgebra
in the deformed $W_\infty$-algebra. They are parametrized by the
complex number $h,\omega,y$ and $h',\omega',y'$. Notice
that their deformation parameters are $ \lambda(h'\omega-h\omega')$.
The previously discussed loop and Yangian subalgebras can be recovered
as particular cases of this construction.

\bigskip
\noindent
{\bf Concluding Remarks.}

We would like to conclude with a few comments.
In this note, we essentially worked with a specific class of
representations of the algebra.
But the algebra can be defined abstractly as the associative algebra
generated by the elements $\{J^{ab}_0,J^{ab}_1,K^{ab}_1\}$ with the
appropriate Serre relations.
So it is important to decipher the statements which are representation
dependent from those which are true in the algebra.
Also we did not discuss the Hopf algebra structure, if any,  of our
algebra.
We hope that this short note not only clarifies the mathematical
structures underlying the relations between the infinite
symmetry~\cite{HiWa93e} and the Yangian symmetry~\cite{HHTBP92,BGHP93}
of the Calogero-Sutherland system, but also suggests various extensions
of the theories.

\bigskip
\par\noindent
{\bf Acknowledgements.}

This short note is a result of discussions on a lecture
note~\cite{WadaHika94a}. It is a pleasure to publish it in
this volume, and the authors take this opportunity to thank
the organizers for their warm welcome. We thank F.~D.~M. Haldane
and V. Pasquier for discussions.

\end{document}